\documentclass[iop]{emulateapj} 
\usepackage{times} 
\usepackage{latexsym,amssymb,amsmath,amsfonts}
\usepackage{rotfloat} 
\usepackage{rotating} 
\usepackage{float} 
\usepackage{graphicx}
\usepackage{longtable}
\usepackage{url}
\usepackage{dcolumn}
\usepackage{placeins}
\usepackage{multirow} 
\usepackage{rotating} 
\begin{document}
\newcommand{\sqcm}{cm$^{-2}$}  
\newcommand{\lya}{Ly$\alpha$}
\newcommand{\lyb}{Ly$\beta$}
\newcommand{\lyg}{Ly$\gamma$}
\newcommand{\lyd}{Ly$\delta$}
\newcommand{\HI}{\mbox{H\,{\sc i}}}
\newcommand{\HII}{\mbox{H\,{\sc ii}}}  
\newcommand{\HeI}{\mbox{He\,{\sc i}}}
\newcommand{\HeII}{\mbox{He\,{\sc ii}}}
\newcommand{\HeIII}{\mbox{He\,{\sc iii}}}  
\newcommand{\OI}{\mbox{O\,{\sc i}}}
\newcommand{\OII}{\mbox{O\,{\sc ii}}}
\newcommand{\OIII}{\mbox{O\,{\sc iii}}}
\newcommand{\OIV}{\mbox{O\,{\sc iv}}}
\newcommand{\OV}{\mbox{O\,{\sc v}}}
\newcommand{\OVI}{\mbox{O\,{\sc vi}}}
\newcommand{\OVII}{\mbox{O\,{\sc vii}}}
\newcommand{\OVIII}{\mbox{O\,{\sc viii}}} 
\newcommand{\CI}{\mbox{C\,{\sc i}}}
\newcommand{\CII}{\mbox{C\,{\sc ii}}}
\newcommand{\CIII}{\mbox{C\,{\sc iii}}}
\newcommand{\CIV}{\mbox{C\,{\sc iv}}}
\newcommand{\CV}{\mbox{C\,{\sc v}}}
\newcommand{\CVI}{\mbox{C\,{\sc vi}}}  
\newcommand{\SiII}{\mbox{Si\,{\sc ii}}}
\newcommand{\SiIII}{\mbox{Si\,{\sc iii}}}
\newcommand{\SiIV}{\mbox{Si\,{\sc iv}}}
\newcommand{\SiXII}{\mbox{Si\,{\sc xii}}}   
\newcommand{\SII}{\mbox{S\,{\sc ii}}}
\newcommand{\SIII}{\mbox{S\,{\sc iii}}}
\newcommand{\SIV}{\mbox{S\,{\sc iv}}}
\newcommand{\SV}{\mbox{S\,{\sc v}}}
\newcommand{\SVI}{\mbox{S\,{\sc vi}}}  
\newcommand{\NI}{\mbox{N\,{\sc i}}}   
\newcommand{\NII}{\mbox{N\,{\sc ii}}}   
\newcommand{\NIII}{\mbox{N\,{\sc iii}}}     
\newcommand{\NIV}{\mbox{N\,{\sc iv}}}   
\newcommand{\NV}{\mbox{N\,{\sc v}}}    
\newcommand{\PV}{\mbox{P\,{\sc v}}} 
\newcommand{\NeIV}{\mbox{Ne\,{\sc iv}}}   
\newcommand{\NeV}{\mbox{Ne\,{\sc v}}}   
\newcommand{\NeVI}{\mbox{Ne\,{\sc vi}}}   
\newcommand{\NeVII}{\mbox{Ne\,{\sc vii}}}   
\newcommand{\NeVIII}{\mbox{Ne\,{\sc viii}}}   
\newcommand{\NeIX}{\mbox{Ne\,{\sc ix}}}   
\newcommand{\NeX}{\mbox{Ne\,{\sc x}}} 
\newcommand{\MgI}{\mbox{Mg\,{\sc i}}}
\newcommand{\MgII}{\mbox{Mg\,{\sc ii}}}  
\newcommand{\MgX}{\mbox{Mg\,{\sc x}}}   
\newcommand{\FeII}{\mbox{Fe\,{\sc ii}}}  
\newcommand{\FeIII}{\mbox{Fe\,{\sc iii}}}   
\newcommand{\NaIX}{\mbox{Na\,{\sc ix}}}   
\newcommand{\ArVIII}{\mbox{Ar\,{\sc viii}}}   
\newcommand{\AlXI}{\mbox{Al\,{\sc xi}}}   
\newcommand{\CaII}{\mbox{Ca\,{\sc ii}}}  
\newcommand{\zabs}{$z_{\rm abs}$}
\newcommand{\zmin}{$z_{\rm min}$}
\newcommand{\zmax}{$z_{\rm max}$}
\newcommand{\zqso}{$z_{\rm qso}$}
\newcommand{\degree}{\ensuremath{^\circ}}
\newcommand{\lapp}{\mbox{\raisebox{-0.3em}{$\stackrel{\textstyle <}{\sim}$}}}
\newcommand{\gapp}{\mbox{\raisebox{-0.3em}{$\stackrel{\textstyle >}{\sim}$}}}
\newcommand{\be}{\begin{equation}}
\newcommand{\en}{\end{equation}}
\newcommand{\di}{\displaystyle}
\def\tworule{\noalign{\medskip\hrule\smallskip\hrule\medskip}} 
\def\onerule{\noalign{\medskip\hrule\medskip}} 
\def\bl{\par\vskip 12pt\noindent}
\def\bll{\par\vskip 24pt\noindent}
\def\blll{\par\vskip 36pt\noindent}
\def\rot{\mathop{\rm rot}\nolimits}
\def\alf{$\alpha$}
\def\refff{\leftskip20pt\parindent-20pt\parskip4pt}
\def\kms{km~s$^{-1}$}
\def\zem{$z_{\rm em}$}

\shortauthors{S. Muzahid}

\title{Probing the large and massive CGM of a galaxy at $z \sim$ 0.2 using a pair of quasars\altaffilmark{1}}   

\author{Sowgat Muzahid\altaffilmark{2}}

\altaffiltext{1}{Based on observations made with the NASA/ESA {\sl Hubble Space Telescope}, obtained from the data archive at the Space Telescope Science Institute, which is operated by the Association of Universities for Research in Astronomy, Inc., under NASA contract NAS 5-26555.}    
\altaffiltext{2}{The Pennsylvania State University, 413 Davey Lab, University Park, State College, PA 16801, USA}       
 

\begin{abstract}  

We present analysis of two \OVI\ absorbers at redshift \zabs\ = 0.227, detected in the spectra of two closely spaced QSO sightlines (Q~0107$-$025 {\bf A} and {\bf B}), observed with the Cosmic Origins Spectrograph (COS) on board the {\it Hubble Space Telescope} ($HST$). At the same redshift, presence of a single bright ($\sim$1.2$L_{\star}$) galaxy at an impact parameter of $\sim$~200 kpc (proper) from both the sightlines was reported by \citet{Crighton10}. Using detailed photoionization models we show that the high ionization phases of both the \OVI\ absorbers have similar ionization conditions (e.g. $\log U \sim$ $-$1.1 to $-$0.9), chemical enrichment (e.g. $\log Z \sim$ $-$1.4 to $-$1.0), total hydrogen column density (e.g. $\log N_{\rm H} (\rm cm^{-2}) \sim 19.6 - 19.7$) and line of sight thickness (e.g. $l_{\rm los} \sim $~600 -- 800~kpc). Therefore we speculate that the \OVI\ absorbers are tracing different parts of same large scale structure, presumably the circumgalactic medium (CGM) of the identified galaxy. Using sizes along and transverse to the line of sight, we estimate the size of the CGM to be $R \sim 330$~kpc. The baryonic mass associated with this large CGM as traced by \OVI\ absorption is $\sim 1.2\times10^{11} M_{\odot}$. A low ionization phase is detected in one of the \OVI\ systems with near solar metallicity ($\log Z = 0.20\pm0.20$) and parsec scale size ($l_{\rm los}\sim$ 6 pc), possibly tracing the neutral phase of a high velocity cloud (HVC) embedded within the CGM.  
 
\end{abstract}  

\maketitle
\section{Introduction} 
\label{sec_intro}  

According to current theoretical models, accretion of pristine gas from the intergalactic medium \citep[IGM;][]{Keres05,Dekel09,Bouche10,Dave12} and efficient galactic-scale outflows of metal-enriched gas \citep[]{Springel03,Oppenheimer10,Dave11a,Dave11b} are the two primary factors that govern the formation and evolution of galaxies. Some of the outflowing metal-enriched material eventually returns to galaxies \citep[]{Oppenheimer10}. This ``baryon cycle" alters the ionization and chemical conditions of the circumgalactic medium (CGM), i.e. gas in the immediate vicinity of galaxies that lies within the dark matter halos. However, due to low density of the CGM gas, direct detection of such a ``baryon cycle" remains a big challenge.       

The absorption lines observed in the spectra of distant quasars (QSOs) allow us to probe this tenuous CGM \citep[]{Tumlinson11Sci,Thom12,Werk13}, which is otherwise not visible to us. However, the major drawback of absorption line spectroscopy is that one does not have any information about variations in the physical conditions of the absorbing gas in the direction transverse to the line of sight. Closely spaced QSO pairs (or groups) can provide useful information regarding transverse size and therefore the tomography of the absorber \citep[]{Bechtold94,Dinshaw95,Dinshaw97,Dodorico98,Rauch01,Petry06,Crighton10}.    

The resonant transitions of five times ionized oxygen (i.e. \OVI$\lambda\lambda$1031,1037) is a useful tracer of low density diffuse gas. The high cosmic abundance of oxygen and high ionization potential (IP = 113.9 eV) provide \OVI\ doublets with immense diagnostic power. \OVI\ absorption is observed in a wide variety of astrophysical environments, e.g. IGM \citep[]{Tripp08,Thom08,Muzahid11,Muzahid12a}, local ISM and Galaxy halo \citep[]{Savage03,Wakker09}, HVC \citep[]{Sembach03}, CGM \citep[]{Tumlinson11,Tumlinson11Sci,Kacprzak12,Stocke13} etc. In particular, in pioneering work, \citet{Tumlinson11Sci} have shown that \OVI\ absorption is ubiquitous in the CGM of isolated star forming galaxies. Moreover, this highly ionized CGM gas contains considerable mass that can account for the ``missing baryons" in galaxies \citep[see also][]{Tripp11}.    

In this paper we present analysis of two \OVI\ absorbers at redshift \zabs\ = 0.227, detected in the spectra of two closely spaced quasars. \citet{Crighton10} have identified a bright galaxy at an impact parameter of $\sim$~200 kpc at the redshift of the absorbers. We use photoionization models to understand the physical conditions of the CGM of the galaxy as probed by the \OVI\ absorption. Moreover, we present a robust estimate of the CGM mass using the model predicted values of metallicity and ionization correction.   
This paper is organized as follows: after presenting observations and data reduction in section~\ref{sec_obs}, data analysis and photoionization models are presented in section~\ref{sec_ana}. In section~\ref{sec_diss} we discuss our results and summarize the conclusions. Throughout this paper we adopt an $H_{0}$ = 70 \kms Mpc$^{-1}$, $\Omega_{\rm M}$ = 0.3 and $\Omega_{\rm \Lambda}$ = 0.7 cosmology. The relative abundance of heavy elements is taken from \citet{Asplund09}. All the distances given are proper (physical) distances.

\begin{figure*} 
\centerline{
\vbox{
\centerline{\hbox{ 
\includegraphics[height=12.9cm,width=12.4cm,angle=00]{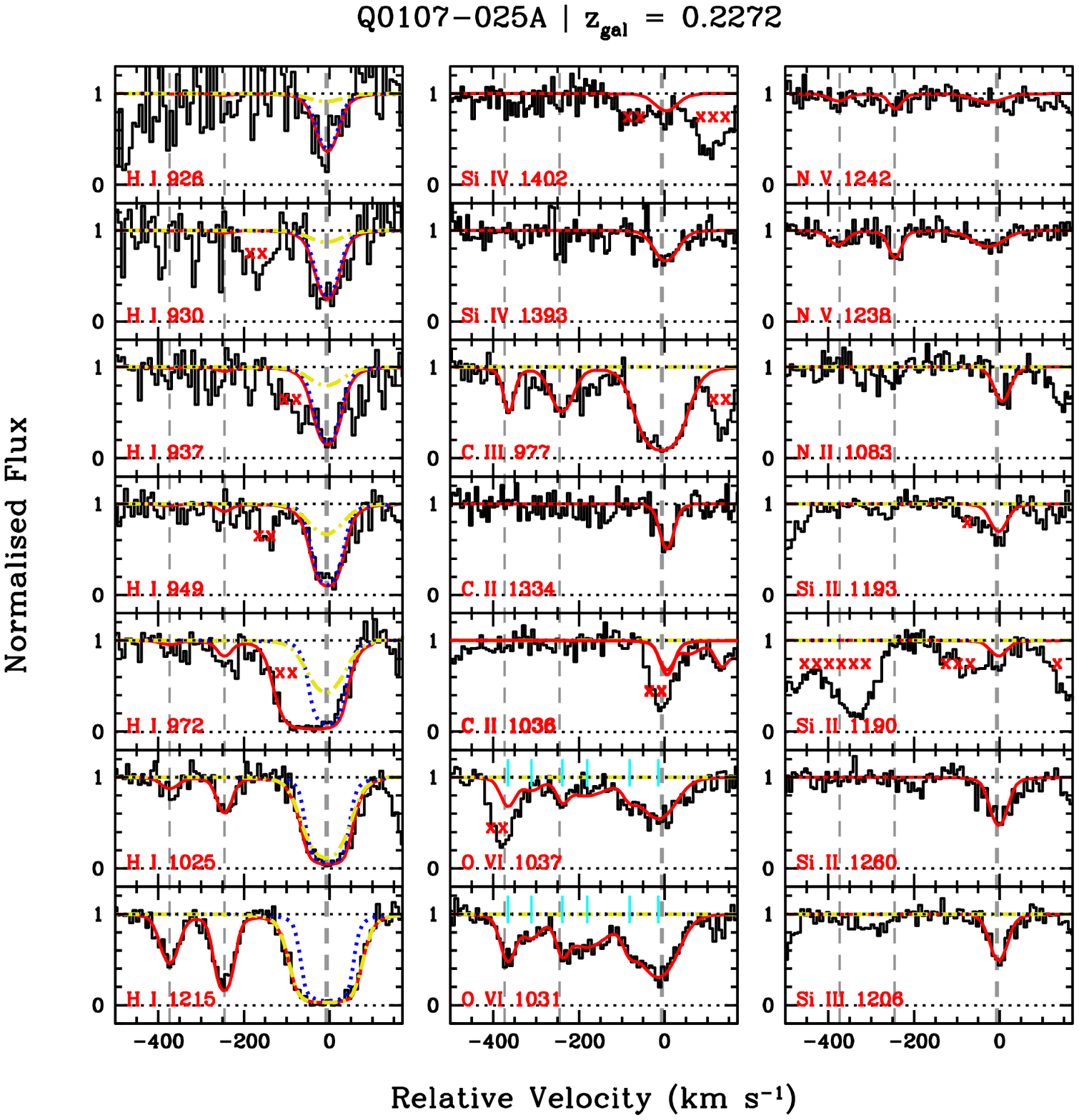}  
\includegraphics[height=12.9cm,width=6.4cm,angle=00]{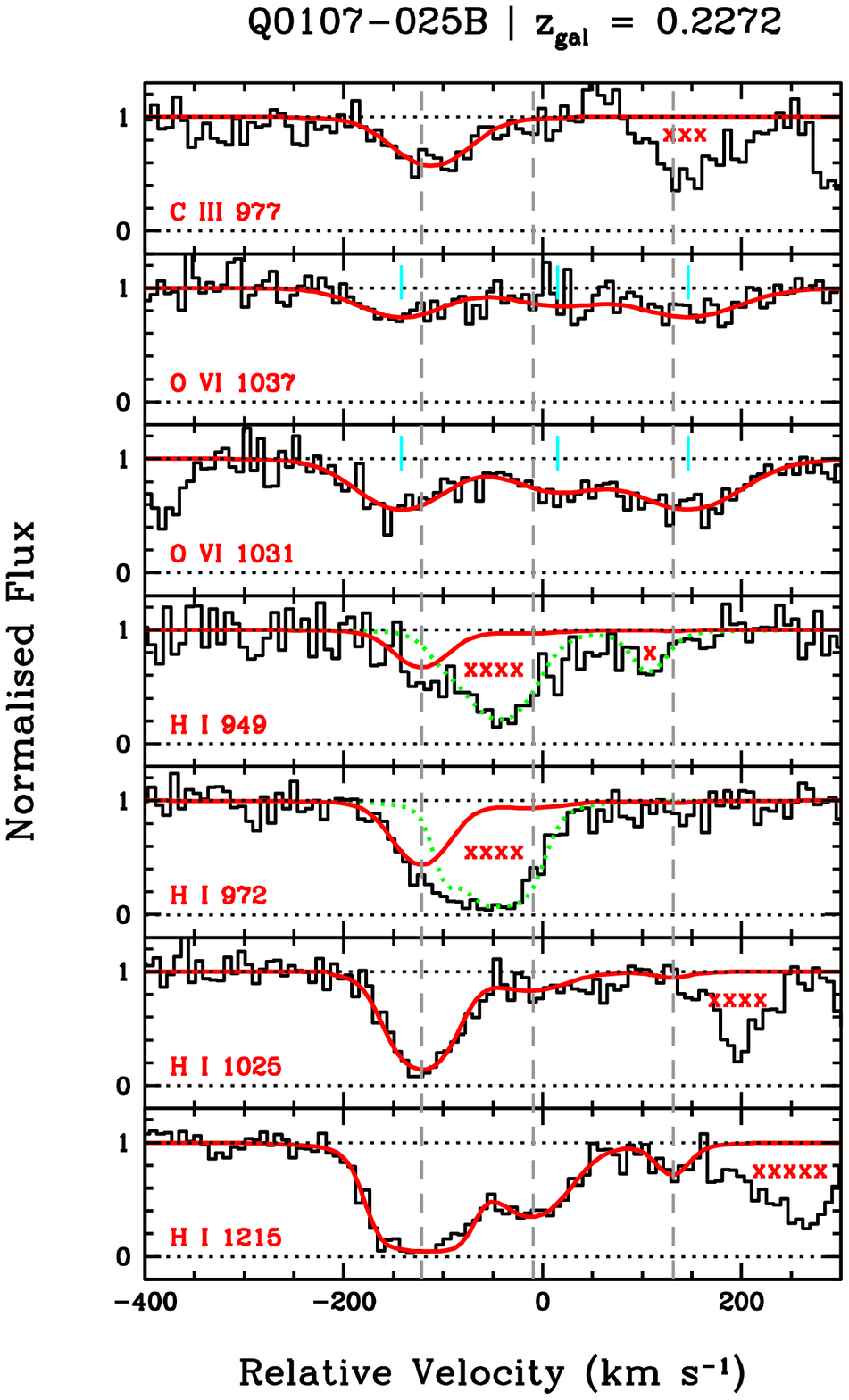}  
}} 
}}  
\caption{Absorption profiles (black histograms) of different species 
in the \zabs\ = 0.227 system, detected in the spectrum of {\bf A} (left) and {\bf B} 
(right), are plotted in velocity. The zero velocity corresponds to the redshift of the galaxy 
$z_{\rm gal}$ = 0.2272. Best fitting Voigt profiles are shown in smooth (red) curves. 
The (blue) dotted and (yellow) dot-dashed curves in the Lyman series lines in the left 
panel show the \HI\ absorption associated with low and high ionization phases respectively. 
The (blue) dotted profile is dominated by the higher order Lyman series absorption whereas the 
(yellow) dot-dashed profile is dominated by the shape of \lya\ and \lyb\ absorption 
(see text). Component centroids of \HI\ absorption are shown by the dashed vertical 
lines. The ticks in the \OVI\ panels mark the line centroids of \OVI\ components. Unrelated 
absorption (blends) are marked by `x'. The (green) dotted profiles in the right panel represent 
model for the blend.}    
\label{vplot} 
\end{figure*} 

\section{Observations, Data reduction and Line measurement}  
\label{sec_obs}

The well known QSO pair (Q~0107$-$025A, \zem\ = 0.960; Q~0107$-$025B, \zem\ = 0.956; hereafter {\bf A} and {\bf B}) were first observed by \citet{Dinshaw95,Dinshaw97} with the $HST$/FOS (Faint Object Spectrograph) on February 1994. \citet{Young01}, subsequently presented observations of another nearby quasar Q~0107$-$0232 (\zem\ = 0.726). Among these three quasars, the minimum angular separation is between the pair {\bf A} and {\bf B} ($\Delta \theta$ = 1.29 arcmin). This angular separation corresponds to a transverse distance of $\sim$ 280 kpc at \zabs\ = 0.227 for our adopted cosmology. The analysis of \lya\ absorbers in this QSO triplet, based on low resolution data, has been presented in several previous papers \citep[e.g.,][]{Dinshaw95,Dinshaw97,Dodorico98,Petry06,Crighton10}. In this paper, we will concentrate on two highly ionized absorbers detected via \OVI\ absorption in medium resolution COS spectra. The third QSO, Q~0107$-$0232, is observed only with the COS/G160M grating and therefore the spectrum does not cover \OVI\ for the redshift of interest. We thus do not use the spectrum of Q~0107$-$0232\ in our analysis.   

The ultraviolet (UV) spectra of Q~0107$-$025 ({\bf A} and {\bf B}) were obtained using $HST/$COS during observation cycle-17, under program ID: 11585 (PI: Neil Crighton). These observations consist of G130M and G160M far-UV (FUV) grating integrations at medium resolution of $ \lambda/\Delta\lambda \sim 20,000$ and signal-to-noise ratio $S/N \sim$ 10 per resolution element in the wavelength range 1134 -- 1796 \AA.
The properties of COS and its in-flight operations are discussed by \cite{Osterman11} and \cite{Green12}. The data were retrieved from the $HST$ archive and reduced using the STScI {\sc CalCOS} v2.17.2 pipeline software. The reduced data were flux calibrated. The alignment and addition of the separate G130M and G160M exposures were done using the software developed by the COS team\footnote{http://casa.colorado.edu/$\sim$danforth/science/cos/costools.html}. The exposures were weighted by the integration time while coadding in flux units. The reduced coadded spectra were binned by 3 pixels as the COS data in general are highly oversampled (6 pixels per resolution element). All measurements and analysis in this work were performed on the binned data. While binning improves the $S/N$ of the data, measurements are found to be fairly independent of binning. Continuum normalization was done by fitting the line free regions with a smooth lower order polynomial. For Voigt profile fit analysis, we use non-Gaussian COS line spread function given by \citet{Kriss11}. Multiple transitions (e.g., doublets or Lyman series lines) were always fitted simultaneously to estimate best fitting column densities.

\section{Analysis} 
\label{sec_ana}

In total, there are 10 \OVI\ systems detected in both {\bf A} and {\bf B}. However, only two systems (i.e. \zabs\ = 0.227 and 0.399), are common to both {\bf A} and {\bf B}. The \zabs\ = 0.227 system is associated with a known bright galaxy \citep[]{Crighton10}. But no galaxy information is available for the other system. In this paper we focus on the \zabs\ = 0.227 system as observed in the spectra of {\bf A} and {\bf B}. Detailed analysis of the other system (i.e. \zabs\ = 0.399) with be presented in an upcoming paper by Muzahid et al. (in preparation).

\subsection{System at $z_{\rm abs} = 0.227$ towards {\bf A} (system {\bf 1A})}   
\label{sec_1A}

Absorption profiles of different species detected in this system are shown in the left panel of Fig.~\ref{vplot}. The \lya\ and \lyb\ absorption clearly show three distinct absorption components spread over $\sim$500 \kms. The two weak \HI\ components (at relative velocities $v_{\rm rel} \sim -$250 and $-$400 \kms) are associated with \CIII, \NV\ and \OVI\ absorption. The strongest \HI\ component at $v_{\rm rel} \sim 0$ \kms\ is associated with low (\NII, \SiII, \CII), intermediate (\CIII, \SiIV) and high (\NV\ and \OVI) ionization metal lines. \citet{Crighton10} report detection of \CIV\ absorption in the $HST/$FOS spectrum with an observed equivalent width of $W_{\rm obs}(1548) = 0.86\pm0.20$ \AA. This corresponds to $\log N(\CIV)$ = 14.24$\pm$0.12, assuming that the line falls on the linear part of the curve of growth. Note that the \SiIV\ absorption is detected only in the \SiIV$\lambda1393$ transition. The other member of the doublet is blended. Moreover, the \SiIV$\lambda1393$ line could have contamination from \NIII$\lambda989$ absorption from the \zabs\ = 0.7286 system. Therefore the measured $N(\SiIV)$ is strictly an upper limit. The \OVI$\lambda1037$ transition in the blue-most component (i.e. at $v_{\rm rel}$$\sim-$400 \kms) is blended with the \lya\ forest. Note that \CII$\lambda1036$ from $v_{\rm rel}$$\sim$ 0 \kms also falls at the same wavelengths. The \OVI\ absorption is spread over $\sim$410 \kms\ and show markedly different profile compared to those of low ions. Minimum six Voigt profile components are required to fit the \OVI\ doublets adequately. To simplify our photoionization model, we summed the \OVI\ column densities of three pairs of nearest-neighbor components (see Table~\ref{fits_tabA}).        

Presence of several unsaturated higher order Lyman series (up to Ly$-926$) lines in the strongest \HI\ component allows robust determination of $N(\HI)$. \lyg\ absorption is severely blended with the Galactic \SiII$\lambda1193$ absorption. We modelled out the contamination while fitting the Lyman series lines. The column density and the Doppler parameter required to fit the higher order lines (i.e. Ly$\delta$, Ly$-937$, Ly$-930$ and Ly$-926$) cannot fully explain the observed \lya\ and \lyb\ absorption. The (blue) dotted profile in the velocity plot (see Fig.~\ref{vplot}) shows the contribution of the component which dominates the profiles of higher order lines. This component has a Doppler parameter of $b(\HI)$ = 26$\pm$3 \kms. It is evident from the \lya\ and \lyb\ profiles that we need another component (preferably broad) to produce a reduced $\chi^2$$\sim$1. The (yellow) dot-dashed profile represents this broad component with $b(\HI)$ = 48$\pm$4 \kms. Because of the presence of multiple ions at different ionization states we focus photoionization model for the component at $v_{\rm rel} \sim0$ \kms. 

\begin{table} 
\begin{center}  
\caption{Column densities in the \zabs\ = 0.227 systems towards {\bf A} and {\bf B}}             
\begin{tabular}{cccr}  
\hline 
{\bf System}  &   \multicolumn {3}{c}{$\log N$ (cm$^{-2}$)} \\ \cline{2-4}
{\bf  1A}     &   \multicolumn {3}{c}{Velocity ($v_{\rm rel}$) range in \kms}  \\ \cline{2-4}         
Species &    ($-$438 to $-$267)       &    ($-$267 to $-$118)   &  ($-$118 to +150)$^{d}$   \\  
\hline \hline  
\HI\ (High)$^{a}$  &   13.53$\pm$0.04   &  14.07$\pm$0.05  & 15.06$\pm$0.19   \\    
\HI\ (Low)$^{b}$   &                    &                  & 15.92$\pm$0.08   \\      
\OVI\              &   14.16$\pm$0.20   &  14.28$\pm$0.28  & 14.57$\pm$0.23   \\   
\NV                &   13.28$\pm$0.11   &  13.48$\pm$0.07  & 13.55$\pm$0.08   \\  
\CIV$^{c}$         &                    &                  & 14.24$\pm$0.12    \\   
\SiIV              &                    &                  & $<$13.18$\pm$0.08   \\  
\CIII              &   13.20$\pm$0.11   &  13.37$\pm$0.05  & 14.20$\pm$0.05   \\   
\SiIII             &                    &                  & 12.90$\pm$0.05   \\   
\CII               &                    &                  & 13.85$\pm$0.07   \\  
\SiII              &                    &                  & 13.04$\pm$0.06   \\  
\NII               &                    &                  & 13.88$\pm$0.08   \\  
\hline \hline 
{\bf System}  &   \multicolumn {3}{c}{$\log N$ (cm$^{-2}$)} \\ \cline{2-4}
{\bf  1B}     &   \multicolumn {3}{c}{Velocity ($v_{\rm rel}$) range in \kms}  \\ \cline{2-4}         
Species &    ($-$243 to $-$56)$^{d}$    &    ($-$56 to $-$74)   &  ($-$74 to +231)   \\  
\hline \hline  
\HI\               &   14.92$\pm$0.05   &  13.82$\pm$0.03  & 13.07$\pm$0.09   \\    
\OVI\              &   14.24$\pm$0.03   &  14.01$\pm$0.06  & 14.30$\pm$0.04   \\   
\CIII              &   13.44$\pm$0.04   &     (blended)    &   (blended)      \\   
\NV                &      $<$ 13.50     &                  &                  \\  
\SiIV              &      $<$ 13.25     &                  &                  \\  
\hline \hline 
\end{tabular} 
\label{fits_tabA}    
Notes -- $^{a}$\HI\ associated with high ionization phase.   
         $^{b}$\HI\ associated with low ionization phase.    
         $^{c}N(\CIV)$ calculated from $W_{\rm obs}$ given in \citet{Crighton10}. 
         $^{d}$photoionization model is done for this clump.     
\end{center}  
\end{table} 

%
\begin{figure*} 
\centerline{
\vbox{
\centerline{\hbox{ 
\includegraphics[height=6.3cm,width=6.2cm,angle=00]{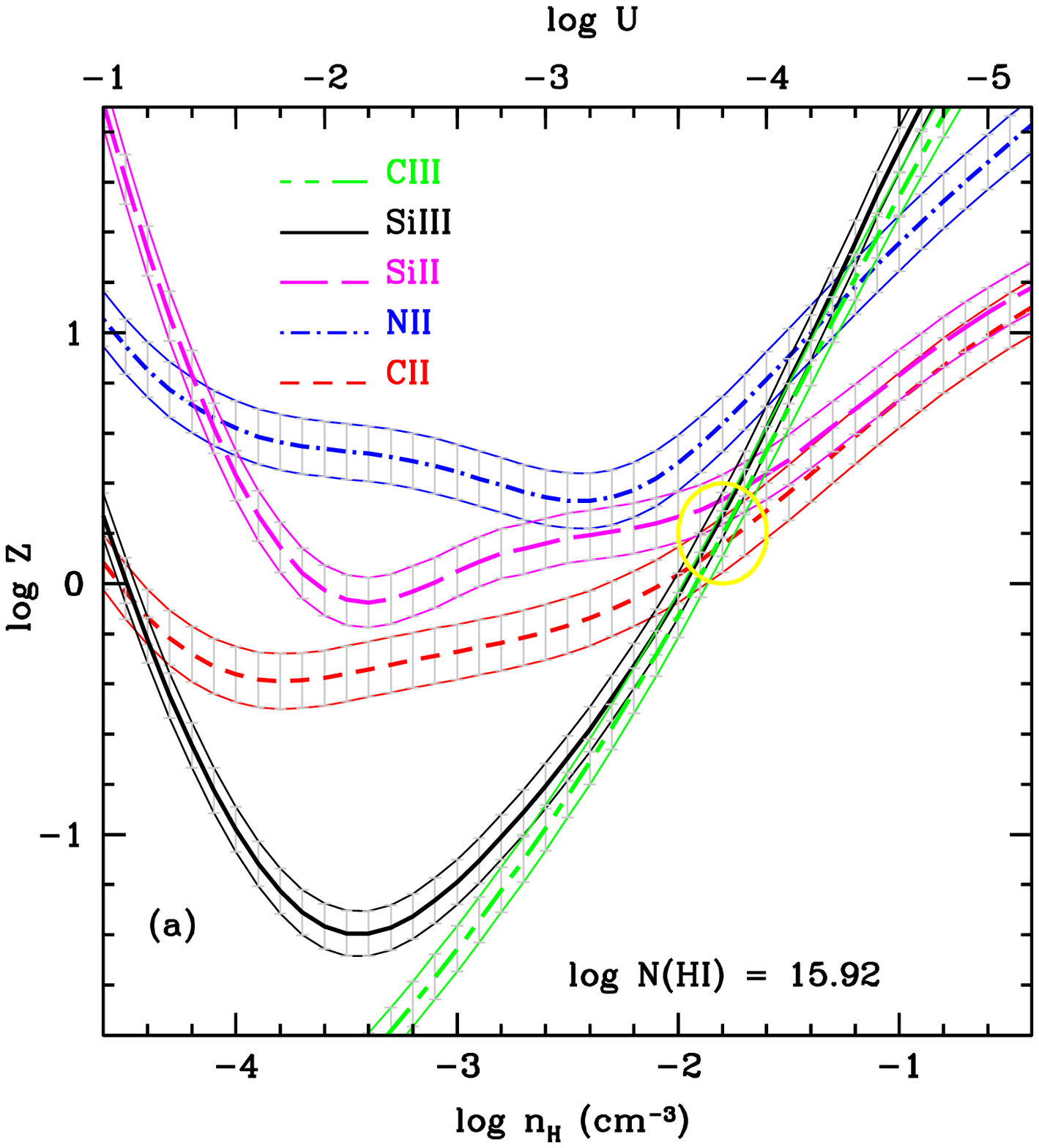}  
\includegraphics[height=6.3cm,width=6.2cm,angle=00]{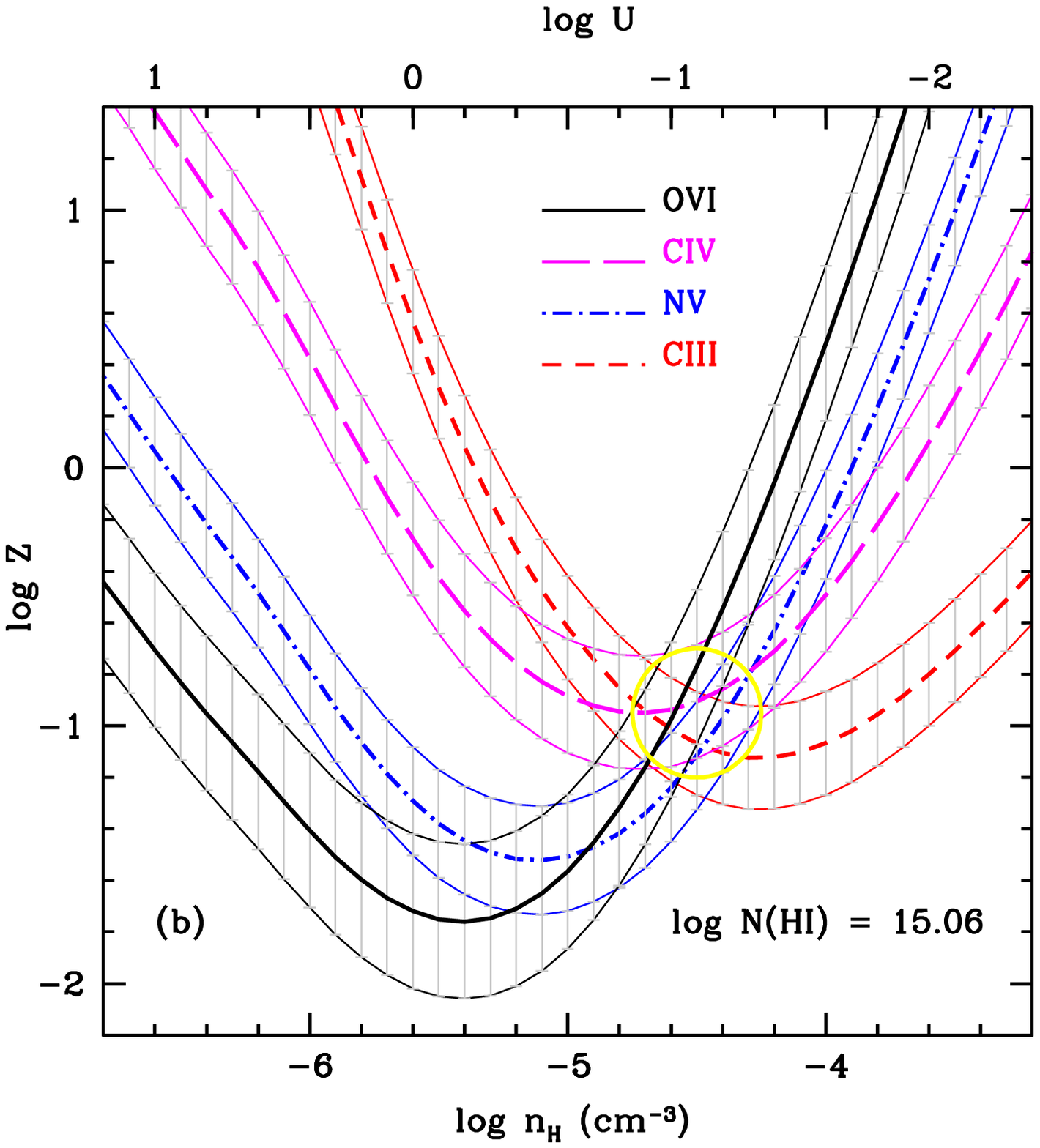}   
\includegraphics[height=6.3cm,width=6.2cm,angle=00]{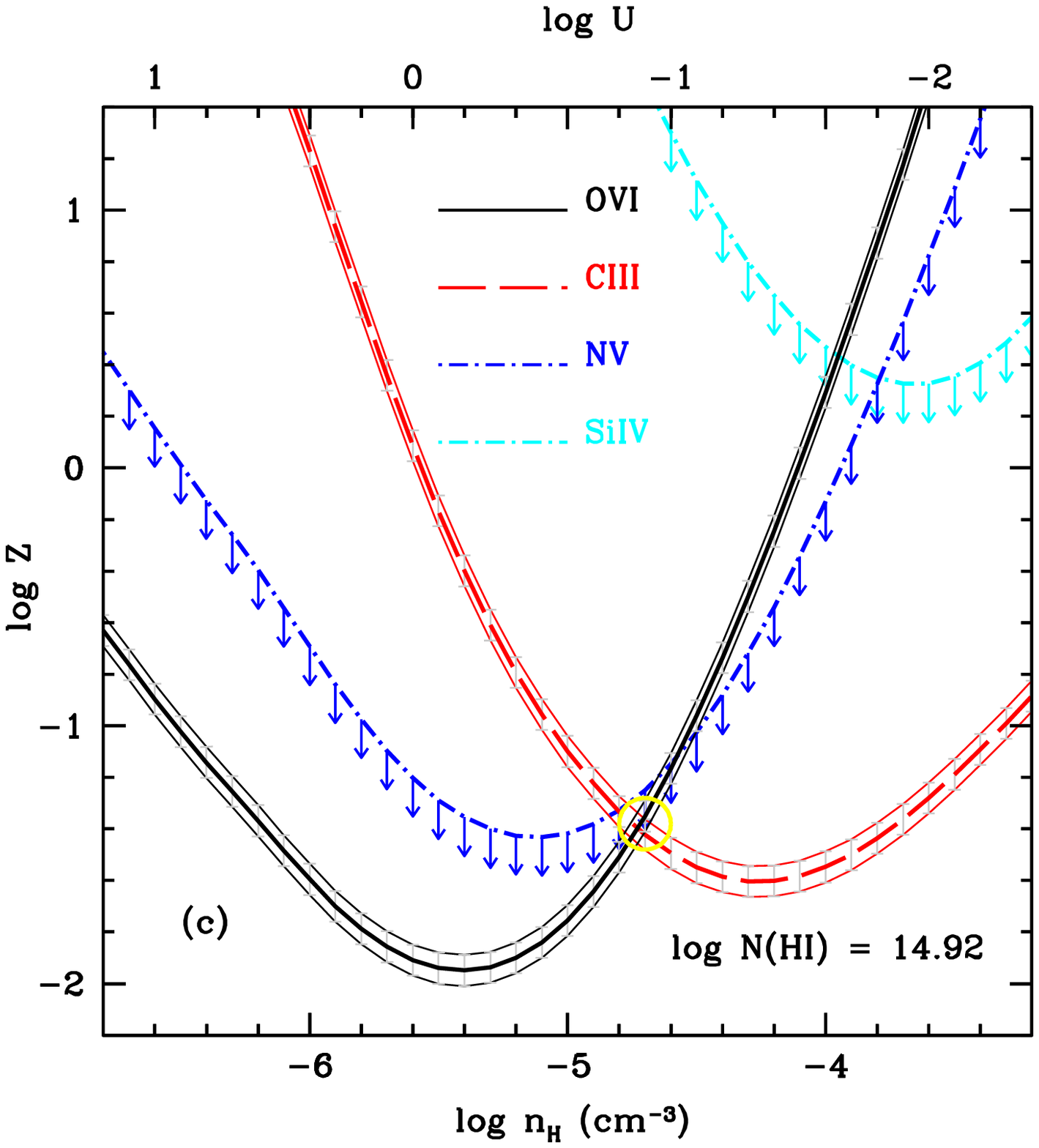}   
}}
}}  
\caption{Photoionization models for the low ionization phase of system {\bf 1A} [panel-(a)], 
high ionization phase of system {\bf 1A} [panel-(b)] and system {\bf 1B} [panel-(c)]. 
Different curves showing loci of different ions in the $\log Z$ -- $\log U$ plane for a given 
$N(\HI)$. The error in each curve is contributed by error in ionic and \HI\ column density 
measurements. The (yellow) circle represents the allowed ranges in $\log U$ and $\log Z$. 
}   
\label{phot_mod}    
\end{figure*} 
%

\subsubsection{Photoionization Model for system {\bf 1A}}   
\label{sec_mod_1A}    

Photoionization models are run using {\sc cloudy} \citep[]{Ferland98} assuming the absorbing gas (a) has plane parallel geometry, (b) has solar relative abundances \citep[]{Asplund09} for heavy elements and (c) is exposed to the extragalactic UV background \citep[]{Haardt96} at redshift $z = 0.23$. Here we do not consider the effect of a galaxy/stellar radiation field, as it is negligible at this redshift at a large separation ($\sim$ 100 kpc) from bright ($> L_{\star}$) galaxies \citep[see e.g.][]{Narayanan10}. We also note that the candidate galaxy does not show any signs of recent star formation \citep[]{Crighton10}.  

First we model the low ionization phase. To constrain the nature of this phase we use column densities of \CII, \SiII, \NII\ and \SiIII. The narrow \HI\ component with higher $N(\HI)$ is associated with this low ionization gas. In panel-(a) of Fig.~\ref{phot_mod} we show the model results computed for $\log N(\HI)$ = 15.92$\pm$0.08. In this plot, different curve represents loci of different low ions in the metallicity -- ionization parameter ($\log Z$ -- $\log U$) plane. The (yellow) circle represents the area in the $\log Z$ -- $\log U$ plane that is allowed by data. The model parameters are summarised in Table~\ref{mod_par}. Note that \NII\ is consistent with this solution only if nitrogen is underabundant by a factor of $\sim$3. This low ionization phase also produce less \NIII\ (i.e. $\log N(\NIII)$ = 12.45) and \CIII\ (e.g. $\log N(\CIII)$ = 13.65) than observed. However, we could not confirm the presence of \NIII\ since the relevant wavelengths are severely affected by Geo-coronal \lya\ emission. Also \CIII\ is an intermediate ion with contribution from both high and low ionization phases. This low ionization phase do not produce any significant high ion absorption, including \SiIV.  

Panel-(b) of Fig.~\ref{phot_mod} show the photoionization model for the high ionization phase of system {\bf 1A}. This phase is constrained by the column densities of \CIII, \CIV, \NV\ and \OVI. Here we use corrected $N(\CIII)$, obtained after dividing out the contribution of the low ionization phase from the \CIII\ profile. We associate the $N(\HI)$ as measured in the broad \HI\ component (see Table~\ref{fits_tabA}) with the high ionization phase. This is a natural choice in view of the Broad \lya\ Absorber (BLA) studies by \citet{Savage11a,Savage11b,Savage12,Narayanan12}, where diffuse gas traced by high ions (e.g. \OVI), is usually associated with a broad \lya\ absorption. In Table~\ref{mod_par} photoionization model parameters are summarized. We notice that within the allowed ranges of $\log U$ and $\log Z$, this high ionization phase can produce significant amount of \NIII, (e.g. $\log N(\NIII)$ = 13.50). But, we could not confirm it because of Geo-coronal \lya\ emission as mentioned earlier. No singly ionized species or \SiIII, are produced in this phase. In passing, we wish to point out that the two weak \HI\ components, at $v_{\rm rel} = -250$ and $-400$ \kms, show ranges in $\log U$ and $\log Z$ which are very similar to this high ionization phase.           

\subsection{System at $z_{\rm abs} = 0.227$ towards {\bf B} (system {\bf 1B})}     
\label{sec_1B}

The velocity plot of this system is shown in the right panel of Fig.~\ref{vplot}. Compared to system {\bf 1A}, this system show weaker \HI\ absorption. No higher order Lyman series lines are detected beyond Ly$\delta$. \lya\ shows three components as also seen in system {\bf 1A}. \lyg\ and \lyd\ lines are blended with the Galactic \SiII$\lambda 1193$ and \OIII$\lambda832$ absorption from \zabs\ = 0.39915 respectively. We have modelled out these blends while fitting. No metal line other than \OVI\ and \CIII\ (only in one component) is detected in this system. The wavelengths redward to the detected \CIII\ is affected by the Galactic \NI$\lambda1199$ line. We note that \OVI\ profile, albeit showing three components as \lya, is not exactly aligned with \lya\ profile. Similar to system {\bf 1A}, the \OVI\ absorption here is also spread over $\sim$405 \kms. \SiIV\ and \NV\ transitions are covered by the COS spectrum, however, we do not detect any measurable absorption at the expected wavelengths. We estimate 3$\sigma$ upper limits on $N(\NV)$ and $N(\SiIV)$ from the observed error spectrum (see Table~\ref{fits_tabA}). Since \CIII\ and \OVI\ are detected only in the blue-most \HI\ component (i.e. at $v_{\rm rel}$$\sim-$120 \kms) we use this component for photoionization modelling.

\subsubsection{Photoionization Model for system {\bf 1B}}    
\label{sec_mod_1B}    

The photoionization model of the component at $v_{\rm rel}$$\sim-$120 \kms\ of this system is shown in panel-(c) of Fig.~\ref{phot_mod}.  We use \HI, \CIII\ and \OVI\ column density measurements and limits on $N(\NV)$ and $N(\SiIV)$ to constrain our model. The model parameters are summarized in Table~\ref{mod_par}. Note that the allowed values of $\log Z$ and $\log U$ are consistent with the non-detections of \NV\ and \SiIV. This single phase solution produces a \CIV\ column density of $\log N(\CIV)$ = 13.67 corresponding to a $W_{\rm obs}$(1548) = 0.19 \AA\ only. This is close to the detection threshold in the FOS/G190H spectrum and therefore consistent with there being no \CIV\ equivalent width reported for this system in \citet{Crighton10}.

%
\begin{table}  
\begin{center}   
\caption{Model parameters in the \zabs\ = 0.227 systems towards {\bf A} and {\bf B}}     
\begin{tabular}{ccrcr}    
\hline 
System      &  $\log U$          &   $\log Z$    &   $\log N({\rm H})$    &   $\log l_{\rm los}^{a}$   \\   
         &                &           &   ($N$ in ${\rm cm^{-2}}$)   &  ($l_{\rm los}$ in kpc)  \\   
\hline \hline  
{\bf 1A}(Low)$^{b}$    &  $-$3.80$\pm$0.20  &      0.20$\pm$0.20  &   17.5$\pm$0.2   & $-$2.2$\pm$0.4  \\   
{\bf 1A}(High)$^{c}$   &  $-$1.10$\pm$0.25  &   $-$0.95$\pm$0.25  &   19.6$\pm$0.3   &    2.6$\pm$0.5  \\    
{\bf 1B}               &  $-$0.90$\pm$0.10  &   $-$1.38$\pm$0.10  &   19.7$\pm$0.1   &    2.9$\pm$0.2  \\ 
\hline \hline  
\end{tabular} 
Notes -- $^{a}$size of the absorber along the line of sight. $^{b}$parameters for low ionization 
         phase. $^{c}$parameters for high ionization phase.    
\label{mod_par}     
\end{center}   
\end{table}   
%

\section{Discussions and Conclusions}   
\label{sec_diss}  

We present analysis of two \OVI\ absorbers (system {\bf 1A} and {\bf 1B}), detected in the spectra of two closely spaced QSOs (Q0107--025{\bf A} and {\bf B}), at redshift \zabs\ = 0.227.  The angular separation between {\bf A} and {\bf B} (1.29 arcmin) corresponds to a transverse separation of $\sim$280 kpc at the absorber's redshift. At the same redshift, a single bright (1.2$L^{\star}$) galaxy at an impact parameter of $\sim$~200 kpc (from both the sightlines {\bf A} and {\bf B}) was identified by \citet{Crighton10}. These authors have measured star formation rate (SFR) $>$~0.45 $\rm M_{\odot} yr^{-1}$ and metallicity $\log Z \ge -$0.30 and claimed that the galaxy did not experience bursts of star formation within the last 2 Gyr. 
Nevertheless, the large velocity spreads ($>$400 \kms) of \OVI\ absorption in both the systems possibly suggest that the highly ionized gas could originate from galactic winds/outflows. But, as there is no recent star formation activity in the candidate galaxy, it cannot be a fresh wind. Such an observation is consistent with the ``ancient outflows" as predicted in a recent simulation by \citet{Ford13}. We note that a wind material moving with a speed of 100 \kms\ can reach a distance of $\sim$200 kpc in 2 Gyr time.   

To understand the physical conditions in the absorbing gas we have built grids of photoionization models. We find that the strongest \HI\ component in system {\bf 1B} can be explained with a single phase photoionization model, whereas, the strongest \HI\ component in system {\bf 1A} required at least two phases to explain all the detected ions. Moreover, we find remarkable similarities between the photoionization model parameters of system {\bf 1B} and the high ionization phase of system {\bf 1A}. For example, they show ionization parameter, $\log U \sim -$1.1 to $-$0.9; metallicity, $\log Z \sim -$1.4 to $-$1.0; total hydrogen column density, $\log N_{\rm H} (\rm cm^{-2}) \sim$ 19.6 -- 19.7 and line of sight thickness, $l_{\rm los} \sim$ 600 -- 800 kpc. All these suggest that the \OVI\ absorption in systems {\bf 1A} and {\bf 1B} are possibly tracing the same large scale structure, presumably the CGM of the galaxy identified at the same redshift.            

\begin{figure} 
\centerline{
\vbox{
\centerline{\hbox{ 
\includegraphics[height=3.5cm,width=3.5cm,angle=00]{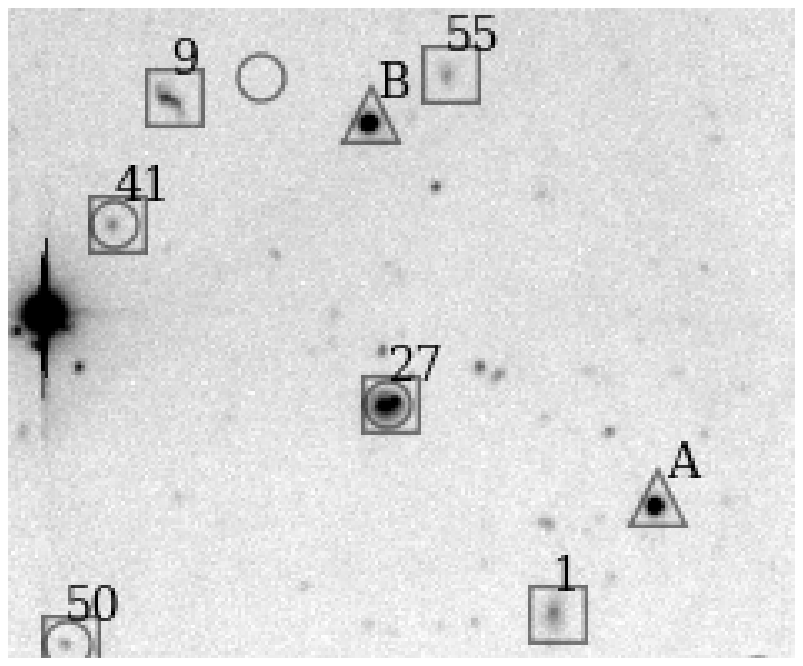}  
\includegraphics[height=3.0cm,width=3.0cm,angle=00]{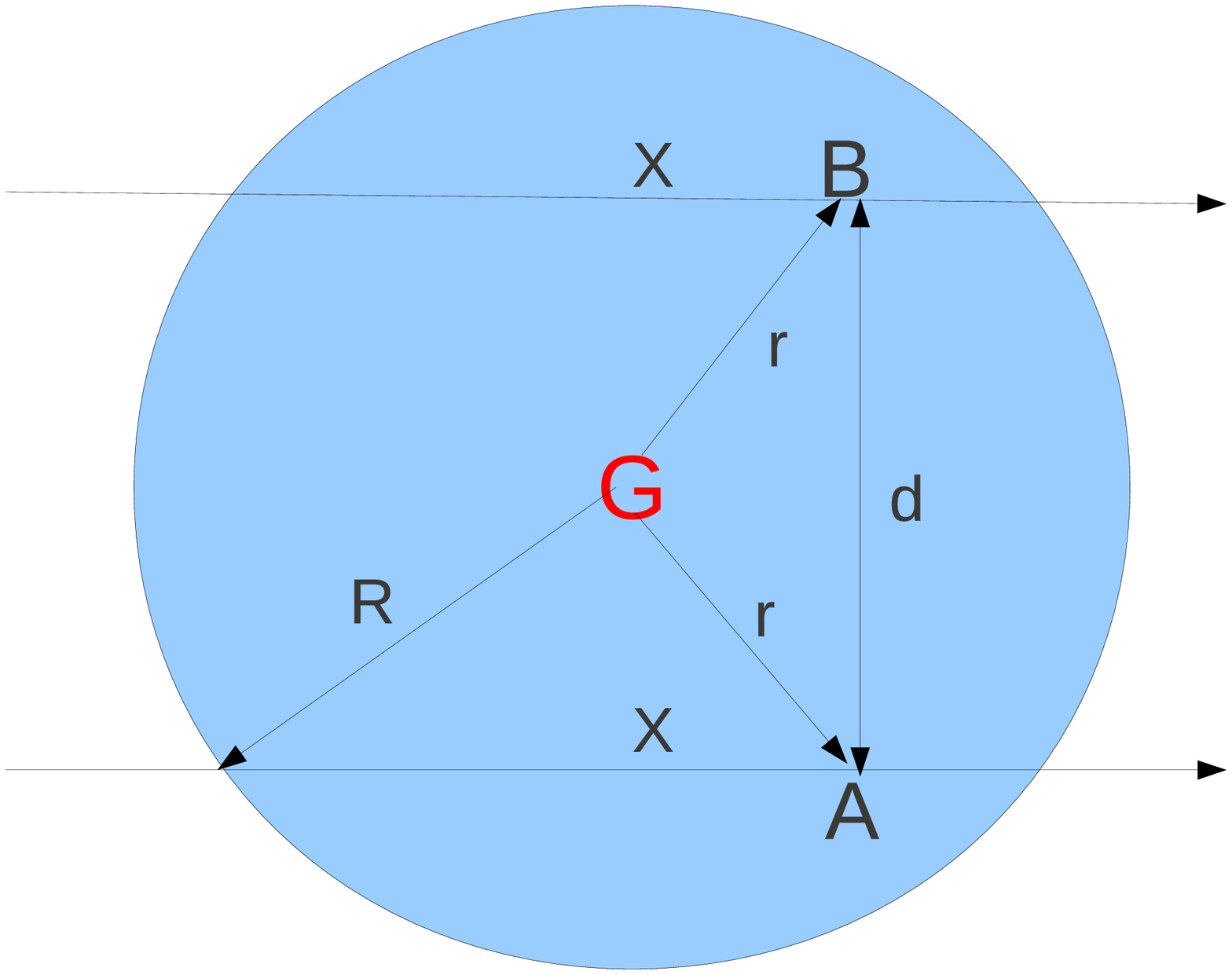}    
}}
}}  
\caption{Left: Zoomed version of Fig.~$1$ of \citet{Crighton10} showing image of the field.  
The box labelled as `27' is the candidate galaxy $\sim$200~kpc away from quasars {\bf A} and 
{\bf B} shown by triangles. The other boxes are galaxies which are not coincident with the 
absorber. Their galaxy sample is complete to an $R$ band magnitude of $\sim$ 21. 
Right: Schematic diagram showing CGM of the galaxy with radius 
$R = \sqrt{(x/2)^{2}+(d/2)^{2}}$. Here $x$ and $d$ are the line of sight thickness ($\sim$ 600 kpc) 
and transverse separation between the absorbers ($\sim$ 280 kpc) respectively.}         
\label{field_plot} 
\vskip1.0cm 
\end{figure}

Fig.~\ref{field_plot} shows the image of the field (left) and a schematic diagram of the CGM (right) as traced by \OVI\ absorption. Using the estimated line of sight thickness and transverse separation between the two absorbers, we estimate size of the CGM, $R \sim$~330 kpc. Assuming the model-predicted density (i.e. $\log n_{\rm H} \sim -$4.6) is uniform inside the sphere of radius $R$, we find CGM mass to be $M_{\rm CGM} \sim 1.2\times 10^{11} {\rm M_{\odot}}$\footnote{Note that the high ionization phase of system {\bf 1A} can also be explained with non-equilibrium collisional ionization models \citep{Gnat07} with $\log T \sim $ 5.2, $\log Z = -1.0$ and $\log N_{\rm H}(\rm cm^{-2})\sim 20.4$, provided nitrogen is underabundant by a factor of $\ge$5 (factor of $\ge$10 in the case of collisional ionization equilibrium). The estimated CGM mass in this case is $\sim$9.8$\times10^{11} {\rm M_{\odot}}$. However, measured Doppler parameters of different species (e.g. $b(\NV)$ = 45$\pm$11 \kms, $b(\OVI)$ = 48$\pm$5 \kms\ and $b(\HI)$ = 48$\pm$4 \kms) do not support such a high temperature.}. Such a large mass in the ionized CGM is also reported by \citet{Tumlinson11Sci,Tripp11}. However, all these studies assume some fiducial values of ionization correction and/or metallicity which are not well constrained by any detailed ionization models.

Our photoionization model suggests that the low ionization phase of system {\bf 1A} can produce $\log N(\MgII) = 12.87$. Therefore this system is a weak \MgII\ absorber candidate. This low ionization phase is compact in size ($l_{\rm los} \sim$ 6 pc) and show high metallicity ($\log Z=$ 0.20$\pm$0.20). Note that the metallicity is $\sim$10 times higher compared to that of the high ionization phase. Such near-solar metallicity and parsec scale size are very common features of weak \MgII\ absorbers \citep[]{Rigby02,Narayanan08,Misawa08}. Moreover, \citet{Narayanan08} have hypothesized that weak \MgII\ absorbers are likely to be tracing gas in the extended halos of galaxies, analogous to the Galactic HVCs. It is intriguing to note that this low ionization phase is detected only in system {\bf 1A} but not seen in system {\bf 1B}. Therefore we put forward a scenario where the low ionization phase traces high density pockets, like HVCs, and the high ionization phase traces extended and diffuse CGM gas. The density difference between the two phases is more than two orders of magnitude. Thus pressure equilibrium would require temperature difference to be of the same order. However, the photoionization temperatures are not that different between the two phases, suggesting that such an absorber could be short-lived. 

%
%

\section{Acknowledgements}       
SM thanks Dr. Anand Narayanan, Dr. Raghunathan Srianand and Dr. Jane Charlton for useful discussions which improved the content significantly.

\bibliographystyle{apj} 

\end{document}